\documentclass[bm,aps,showpacs,nofootinbib,amsfonts,amssymb,twocolumn,preprintnumbers]{revtex4}

\usepackage[dvipdfmx]{graphicx}
\usepackage{amssymb,amsfonts,amsmath,bm,color,multirow,cases,empheq,hyperref}
\usepackage{mathrsfs}

\usepackage{enumerate}
\usepackage{ulem}
\usepackage{afterpage}
\hypersetup{%
 setpagesize=false,
 bookmarksnumbered=true,%
 bookmarksopen=true,%
 colorlinks=true,%
 linkcolor=blue,%
 citecolor=red}
\if0

\usepackage{amsmath,amssymb}
\usepackage{graphicx,enumerate,color}

\fi

\newcommand{\fr}[1]{\frac{1}{#1}}

\newcommand{\ord}[1]{{\mathcal O}(#1)}

\newcommand{\cA}{{\mathcal A}}

\newcommand{\cR}{{\mathcal R}}
\newcommand{\cL}{{\mathcal L}}

\newcommand{\nonum}{\nonumber\\ }

\newcommand{\sR}{{\sf R}}

\newcommand{\cout}[1]{}
\preprint{TTI-MATHPHYS-11}

\begin{document}

\title{Rotating black holes at large $D$ in Einstein-Gauss-Bonnet theory }
\author{Ryotaku Suzuki}
\email{sryotaku@toyota-ti.ac.jp}
\author{Shinya Tomizawa}
\email{tomizawa@toyota-ti.ac.jp}
\affiliation{Mathematical Physics Laboratory Toyota Technological Institute\\
Hisakata 2-12-1, Nagoya 468-8511, Japan}
\date{\today}

\begin{abstract}
Applying the large $D$ approach to the Einstein-Gauss-Bonnet theory, we construct equally rotating black hole solutions in odd dimensions.
This provides the first example of the analytic solutions which describes not-slowly rotating black holes.
For the next-leading order solutions in the $1/D$ expansion, we discuss the physical aspects such as thermodynamics and the phase diagram.
\end{abstract}

\maketitle
The Einstein-Gauss-Bonnet (EGB) theory is a simplest extension of the Einstein theory to the theory with higher curvature terms, which describes string theory inspired ultraviolet corrections to the Einstein gravity~\cite{Gross:1986iv}. 
In particular, the EGB theory in $D=5$ can be regarded as the low energy limit of string theory when the theory is dimensionally reduced from $D=11$ to $D=5$ by compactifying six of the eleven dimensions in compact Calabi-Yau threefold~\cite{Antoniadis:1997eg,Ferrara:1996hh}. 
Furthermore, such quadratic terms of curvatures appears as a $1$-loop correction of heterotic string theory~\cite{Gross:1986iv}.
 Thus, the physics of black holes in the $D=5$ EGB theory has  been the subject of increased attention from the reason that it provides us some insight on a quantum aspect of black holes.

\medskip
The first exact solutions of  black holes in the EGB theory were found by Boulware and Deser for a spherically symmetric and static case in Ref.~\cite{Boulware:1985wk}. 
The static solutions were also generalized to an electrically charged case~\cite{Wiltshire:1988uq,Wiltshire:1985us}.
However, so far, finding rotating black hole solutions in the EGB theory has been considered to be a hard and unsolved problem, since the Kerr-Schild formalism which is a powerful tool for finding rotating black hole solutions cannot work at all  in this EGB theory. 
In spite of the technical difficulty, there are some attempts to construct rotating EGB black hole solutions. 
%For an exact solution was found for the Chern-Simon gravity in $D=5$, but the solution does not have a horizon. 
Equally rotating black hole solutions in $D=5$ were obtained as numerical solutions~\cite{Brihaye:2008kh}, and slowly rotating charged  AdS black hole solutions in $D\ge5$ were obtained  as perturbative and analytic solutions~\cite{Kim:2007iw}. 
 
 \medskip
The large dimension limit or, large $D$ limit~\cite{Asnin:2007rw,Emparan:2013moa,Emparan:2020inr} is a useful approximation, which largely simplifies the black hole analysis in higher dimensions. %without taking scaling limits other than $D\to\infty$. 
Because of the localization of the gravity at large $D$, the dynamical degrees of freedom of the horizon are confined within a thin layer of near-horizon region, which form an effective theory insensitive to the global structure of the spacetime~\cite{Emparan:2015hwa,Bhattacharyya:2015dva,Bhattacharyya:2015fdk,Emparan:2015gva}. 

 \medskip
So far the large $D$ effective theory approach has been a viable tool to study the black hole dynamics not only in general relativity (GR), but also in the EGB theory. 
The (in)stabilities of the static EGB black holes~\cite{Chen:2017hwm}
and black strings~\cite{Chen:2017rxa} were studied by using the large $D$ approach, in which the black string instability is weakened by the Gauss-Bonnet (GB) term for the small GB coupling, whereas enhanced for the large GB coupling. 
Moreover, black ring solutions at large D in the EGB theory were also studied~\cite{Chen:2018vbv}, where they obtained the quasi-normal modes of the EGB black ring and showed that the thin EGB black ring becomes unstable against non-axisymmetric perturbation.
 \medskip

In this letter, we construct new rotating black hole solutions with equal angular momenta in an odd dimensional EGB theory by using the $1/D$-expansion up to the next-to-leading order (NLO).  
The assumption of equal angular momenta in odd dimensions enhances a spacetime symmetry to a class of cohomogeneity one.
The further key assumption is that the metric of a rotating black hole at $D\to\infty$ is locally similar to that of the boosted black string, 
which  was first noticed in the studies of rotating black holes in GR~\cite{Emparan:2013xia,Emparan:2014jca}.  
By imposing this assumption, the leading order equations are decoupled to be simply solvable.
The thermodynamic property is also studied up to the relevant order in $1/D$.

 \medskip
The action of the EGB theory is given by
\begin{align}
 S_{\rm EGB} = \fr{16\pi G} \int \sqrt{-g} \left(R+\alpha_{\rm GB} \cL_{\rm GB} \right)d^D x, 
\end{align}
where the GB Lagrangian is
\begin{align}
\cL_{\rm GB} = R^2-4R_{\mu\nu}R^{\mu\nu}+R_{\mu\nu\rho\sigma}R^{\mu\nu\rho\sigma}.
\end{align}
The equations of motion become
\begin{align}
 R_{\mu\nu}+\fr{2}Rg_{\mu\nu} + \alpha_{\rm GB} H_{\mu\nu}=0,\label{eq:EGB-eom}
\end{align}
where
\begin{align}
&H_{\mu\nu} = -\fr{2}\cL_{\rm GB}g_{\mu\nu}+2RR_{\mu\nu}-4R_{\mu\alpha}R^\alpha{}_\nu \nonum
&\hspace{1.2cm}- 4 R_{\mu\alpha\nu\beta}R^{\alpha\beta} + 2 R_{\mu\alpha\beta\gamma}R_{\nu}{}^{\alpha\beta\gamma}.
\end{align}
The outcome of the large $D$ limit depends on which scales are fixed in the limit, i.e., which scale of physics we are going to focus on. To obtain the black hole horizon, we must fix the length scale of the horizon radius $r_0$ at $\ord{1}$. %, and hence we set $r_0=1$ later. 
 With the fixed horizon scale $r_0$, the scalar curvature around the horizon has the magnitude of $\ord{D^2/r_0^2}$. We are interested in the intermediate regime in which the Einstein-Hilbert and GB terms become comparable $R\sim \alpha_{\rm GB}\cL_{\rm GB}\sim \alpha_{\rm GB} R^2$, otherwise the equation of motion reduces to that of the Einstein or pure GB theory. Thus, we assume the GB coupling scales as $\alpha_{\rm GB} =\ord{r_0^2/D^{2}}$ at large $D$.

 \medskip
Even for the large $D$ limit, it is not so easy to solve the Einstein equations under the general rotating ansatz since the metric functions are non-linearly coupled already at the leading order. 
Instead, we assume that the EGB rotating black holes have the same property as GR rotating black holes, i.e., the large $D$ limit of the Myers-Perry metric reduces to that of the boosted black brane~\cite{Emparan:2013xia,Emparan:2014jca}.
%the following property in GR \red{rotating} black holes also \sout{applies to} \red{holds for}, i.e., 
%that the large $D$ limit of the Myers-Perry metric reduces to that of the boosted black brane\sout{, which can be seen as the static black brane formally in the boosted frame}~\cite{Emparan:2013xia,Emparan:2014jca}.
For instance, in the Einstein-Maxwell theory, the same strategy has been successful in constructing charged rotating black holes in the large $D$ limit both with a single angular momentum~\cite{Mandlik:2018wnw} and equal angular momenta~\cite{Tanabe:2016opw}.

 \medskip
We thus start from the following metric ansatz of equally rotating black holes in $D=2n+3$ dimensions with the Eddington-Finkelstein gauge
\begin{align}
 &ds^2 = -A(r) (e^{(0)})^2+2U(r) e^{(0)} e^{(1)} + 2C(r) e^{(0)}e^{(2)}\nonum
&  \hspace{4cm}+ H(r)(e^{(2)})^2+r^2 d\Sigma^2,
\label{eq:ansatz}
\end{align}
where $d\Sigma^2$ is the Fubini-Study metric on $CP^n$
and other tetrad bases are defined by
\begin{align}
&e^{(0)} = \frac{dt-\Omega r(d\phi+\cA)}{\sqrt{1-\Omega^2}},\quad
  e^{(2)}= \frac{r(d\phi+\cA)-\Omega dt}{\sqrt{1-\Omega^2}},\nonum
&  e^{(1)} = dr,
\quad
\end{align}
with the dimensionless spin parameter $\Omega$ which produces the local Lorentz boost in the subspace $(dt,r(d\phi+\cA))$. Here $\cA$ is the K\"ahler potential of $CP^n$. 
In what follows, we use $1/n$ as the expansion parameter rather than $1/D$ itself, since the large $D$ owes to the large dimension of $CP^n$.
%To impose the metric resembling the static form in the boosted frame at $n\to\infty$, we assume
We impose that the metric reduces to that of the boosted black brane at $n\to\infty$
~\footnote{The assumption $C={\cal O}(n^{-1})$ alone gives $H,U={\rm const}+\ord{n^{-1}}$ in GR. However, we could not decouple the leading order equation only with the assumption for $C$ in the EGB theory.}
\begin{align}
C =\ord{n^{-1}},\quad H = 1+\ord{n^{-1}}.\label{eq:assume-bstbrane}
\end{align}
As the asymptotic boundary condition, we impose
\begin{align}
 A \to 1,\quad U \to 1 ,\quad C \to 0,\quad H\to 1
\end{align}
at $r \to \infty$, so that the ansatz~(\ref{eq:ansatz})
is asymptotically flat.
To resolve the thin near-horizon region at the large $n$ limit, we introduce the following often-used radial coordinate
\begin{align}
  \sR:=r^{2n}.
\end{align}
Here we set the horizon scale $r_0=1$ using the scaling degree of freedom.
The metric components are expanded by $1/n$ as a function of $\sR$
\begin{align}
& A = \sum_{i=0}^\infty \fr{n^{i}} A_i(\sR), \quad U = \sum_{i=0}^\infty  \fr{n^{i}}U_i(\sR), \nonum
 & C = \sum_{i=0}^\infty  \fr{n^{i}} C_i(\sR),\quad  H= \sum_{i=0}^\infty \fr{n^{i}}H_i(\sR).
\end{align}
To keep the Einstein-Hilbert and GB terms comparable at the large $n$ limit in eq.~(\ref{eq:EGB-eom}), we introduce the rescaled GB coupling parameter which remains finite at $n\to\infty$,
\begin{align}
\alpha := (2n)^2 \alpha_{\rm GB}. 
\end{align}
%\subsection{Leading order}
With the assumption~(\ref{eq:assume-bstbrane}),
we can decouple the leading order equation, which yields
\begin{align}
& A_0 = 1+\fr{2\alpha} - \fr{2\alpha}\sqrt{1+\frac{4\alpha(\alpha+1)m}{\sR}},\nonum
&U_0 = 1,\quad C_0=0,\quad H_0=1, \label{eq:LO-sol}
\end{align}
where the integration constant $m$ introduces the horizon at $\sR=m$. 
As one can see in the form of $A_0$, the leading order metric, therefore, reduces to the boosted black string metric at large $D$ as in GR~\cite{Chen:2017rxa}. Note that, for the existence of the horizon, we only consider the parameter region $\alpha > -1/2$.
%\subsection{Next-to-leading order}

\medskip
To obtain the information for $D<\infty$,
we need to solve the $1/n$ correction to the above leading order metric.
In the higher order analysis, $A_i$ and $C_i$ get extra integration constants, which are not determined by the boundary condition. They actually correspond to the parameter shift in the mass parameter $m$ and horizon velocity $\Omega_H$ in each order of $n^{-i}$. Here $\Omega_H$ is determined so that $k = \partial_t + \Omega_H\partial_\phi$ becomes the null generator of the horizon.
To fix the above integration constants, we simply set
\begin{align}
 A_i(\sR=m)=0,\quad C_i(\sR=m)=0.
\end{align}
This sets the horizon at $\sR=m$ and 
angular velocity as
\begin{align}
 \Omega_H = \Omega\, m^{-\fr{2n}}\label{eq:omega_H}
\end{align}
 in all order of $1/n$.
In the original coordinate, the horizon radius is given by
\begin{align}
 r_H:= m^\fr{2n}.
\end{align}
For the other metric functions, we simply impose the regularity at $\sR=m$ and asymptotic boundary condition.
In the derivation, it is convenient to introduce an auxiliary variable~\cite{Chen:2018vbv}
\begin{align}
X:=\sqrt{1+\frac{4\alpha(\alpha+1)m}{\sR}},
\end{align}
which takes $X=1$ at $\sR =\infty$ and $X=1+2\alpha$ on the horizon.

Having these in mind, the next-to-leading order solution is determined as
\begin{align}
& C_1 = \frac{\Omega(X-1)}{4\alpha(1-\Omega^2)}\log \left(\frac{4\alpha(1+\alpha)}{X^2-1}\right),
\end{align}
\begin{align}
&U_1 = \frac{(X-1) \Omega ^2 (\alpha  (X-1)-1)}{2 (\alpha +1) (2 \alpha +1)
   \left(X^2+1\right) \left(\Omega ^2-1\right)},
\end{align}
\begin{align}
& H_1=\frac{\Omega ^2}{ (\alpha +1) (1-\Omega ^2)} \left[  \log \left(\frac{X+1}{2}\right)-\arctan X\right.\nonum
 & \hspace{2cm} \left.+\,\frac{\pi}{4}- \fr{2(2\alpha+1)}\log \left(\frac{X^2+1}{2}\right)\right],
\end{align}
and
\begin{widetext}
\begin{align}
&A_1 = \frac{\left(X^2-1\right) \Omega ^2 \log \left(X^2+1\right)}{16 \alpha  \left(2 \alpha ^2+3
   \alpha +1\right) X \left(\Omega ^2-1\right)}+\frac{\left(X^2-1\right) \Omega ^2 (\arctan
   X-\arctan(1+2\alpha))}{8 \alpha  (\alpha +1) X \left(\Omega ^2-1\right)}
   -\frac{(X-1) \left(X+2 \Omega
   ^2-1\right) \log (X-1)}{4 \alpha  X \left(\Omega ^2-1\right)}
   \nonum
   &\hspace{3cm}-\frac{(X-1) \log (X+1)
   \left(\alpha  \left(4 \Omega ^2-2\right)+X \left(2 \alpha +\Omega ^2+2\right)+5 \Omega
   ^2-2\right)}{8 \alpha  (\alpha +1) X \left(\Omega ^2-1\right)}+a_0+a_1 X+\frac{a_2}{X}, 
   \end{align}
\end{widetext}
\newpage
where the coefficients $a_0,a_1,a_2$ are given by
\begin{align}
&a_0=\frac{\log (4m\alpha (\alpha +1))}{2 \alpha } + \fr{4\alpha(1-\Omega^2)},\\
&a_1=\frac{2(\alpha+1)\log(4 \alpha (1+\alpha))+2\log m+1}{8 \alpha(1+\alpha)(\Omega^2-1)}\nonum
& + \Omega^2 \frac{2(1+2\alpha)\log(2(1+\alpha)/m^2)-\log(1+(1+2\alpha)^2)}{16\alpha(1+\alpha)(1+2\alpha)(\Omega^2-1)},\\
& a_2 = \frac{(1+2\alpha)(1+2\log m)}{8\alpha(1+\alpha)(\Omega^2-1)}+\frac{\log(4\alpha(1+\alpha))}{4\alpha(\Omega^2-1)}\nonum
& -\frac{\Omega ^2 (4 (\alpha +1) \log (2\alpha) 
+(4 \alpha+5)  \log (2(\alpha +1))
)}{8 \alpha  (\alpha +1)(\Omega^2-1)}\nonum
& + \frac{\Omega^2(\log(1+(1+2\alpha)^2)-4(2\alpha+1)^2\log m)}{16\alpha(1+\alpha)(1+2\alpha)(\Omega^2-1)}.
\end{align}
One can easily check that the GR limit $\alpha\to 0$ reproduces the equally rotating Myers-Perry solutions up to NLO in the corresponding gauge.
The large $\alpha$ limit gives another simplification
\begin{align}
& A \to 1 - \sqrt{\frac{m}{\sR}}
 \left(1+\frac{m\log(\sR/m)}{2n\sR(1-\Omega^2)}\right),
\end{align}
\begin{align}
&   C \to \sqrt{\frac{m}{\sR}}\frac{\Omega\log(\sR/m)}{2n(1-\Omega^2)},
  \end{align}
  and
 \begin{align}
 U\to  1+\ord{n^{-2}},\quad H \to 1+\ord{n^{-2}}.
\end{align}
which could imply the existence of the analytic form in the pure GB theory.

\medskip
The ergosurface of the leading order metric~(\ref{eq:LO-sol}) is given by the same condition as in GR
\begin{align}
0=g_{tt} = (1-\Omega^2)^{-1}(-A-2\Omega C+\Omega^2 H),
\end{align}
which is solved as
\begin{align}
 \sR_{\rm ergo} = \frac{(1+\alpha)m}{(1-\Omega^2)(1+\alpha(1-\Omega^2))}+\ord{n^{-1}}.
\end{align}
This is a monotonically increasing function of $\alpha$,
and hence the ergoregion is extended by the GB correction.
For $\alpha\to\infty$, $\sR_{\rm ergo}$ approaches to a finite value.

\medskip
%\section{Thermodynamics}
In the EGB theory, the thermodynamic variables are obtained as in GR, except the entropy defined by the Iyer-Wald formula~\cite{Wald:1993nt,Iyer:1994ys}
\begin{align}
 {\cal S} =\fr{4G} \int_H (1+2\alpha_{\rm GB} \cR) \sqrt{h}\, d^{D-2}x,
\end{align}
where $h$ and $\cR$ is the spacial metric and curvature of the horizon cross section, respectively. Note that the angular velocity is already given in eq.~(\ref{eq:omega_H}).
Up to NLO, the ADM mass and angular momentum, temperature and entropy are given by
\begin{widetext}
\begin{align}
& M =\frac{n\Omega_{2n+1}}{8 \pi G} \frac{(1+\alpha_H)m}{1-\Omega^2}\left[1-\fr{8 n \left(1-\Omega ^2\right) \left(\alpha _H+1\right) \left(2 \alpha _H+1\right)}\left(4-8 \Omega ^2 \alpha _H^2+\left(-8 \Omega ^4+2 (\pi -6) \Omega ^2+8\right) \alpha _H\right.\right.\nonum
&\left.\left. \quad -2 \Omega
   ^2 \log \left(2 \alpha_H^2+2\alpha_H+1\right)
   +4 \Omega ^2 \left(2 \alpha
   _H+1\right) (\log \left(\alpha _H+1\right)-\arctan\left(2 \alpha _H+1\right))+ \Omega ^2(\pi -4)\right)\right],\\
& J =\frac{n\Omega_{2n+1}}{8 \pi G} \frac{(1+\alpha_H)m^\frac{2n+1}{2n}\Omega}{1-\Omega^2}\left[1-\fr{8 n \left(1-\Omega ^2\right) \left(\alpha _H+1\right) \left(2 \alpha _H+1\right)}\left(8 \left(2 \Omega ^2-1\right) \alpha _H^2
+2 \Omega
   ^2 \log \left(2 \alpha _H^2+2 \alpha _H+1\right)\right.\right.\nonum
&\left.\left. \quad  +4(1+2\alpha_H) \Omega ^2 (\arctan\left(2 \alpha   _H+1\right)-\log(\alpha_H+1))-2 \alpha _H \left((\pi -16) \Omega ^2+10\right)
   -(\pi-8)  \Omega ^2-8\right)\right],\\
 & T =\frac{n}{\pi} \frac{1+\alpha_H}{1+2\alpha_H}m^{-\fr{2n}}\sqrt{1-\Omega^2}\left[1-\frac{\left(4 \Omega ^2+1\right) \alpha _H+4 \alpha _H^2+2 \Omega ^2}{2 n \left(1-\Omega
   ^2\right) \left(\alpha _H+1\right) \left(2 \alpha _H+1\right)}\right]\label{eq:temp},\\
& S = \frac{\Omega_{2n+1}}{4G}\frac{(1+2\alpha_H)m^\frac{2n+1}{2n}}{\sqrt{1-\Omega^2}}\left[1+\fr{8 n (1-\Omega^2 )  \left(\alpha _H+1\right) \left(2 \alpha _H+1\right)}\left(8\left(1-2\Omega ^2\right) \alpha _H^2+8 \alpha _H \left(1-2 \Omega ^2\right)
\right.\right.\nonum
   &\left.\left. \qquad+ 4(1+2\alpha_H)\Omega ^2\left( \log(1+\alpha_H)-\arctan\left(2 \alpha _H+1\right)+\pi/4  \right)-2 \Omega ^2    \log \left(2 \alpha _H^2+2 \alpha _H+1\right)\right)\right].
\end{align}
\end{widetext}
where the GB coupling is written in the scale invariant form $\alpha_H := \alpha/r_H^2=\alpha/m^\fr{n}$.
The first law $dM = T dS + \Omega_H dJ$
is easily checked by differentiating with $m$ and $\Omega$ up to NLO with $\alpha$ fixed.

\medskip
From eq.~(\ref{eq:temp}), one can expect the extremal limit would exist approximately at
\begin{align}
\Omega = 1 -\frac{2+5\alpha+4\alpha^2}{4n(1+2\alpha)(1+\alpha)}.
\end{align}
Unfortunately, we will see that $T$ includes $(1-\Omega^2)^{-2}$ in NNLO~\cite{STnext}, which invalidates the $1/n$-expansion around the extremal limit. 
This fact should not be so remarkable, since as pointed out already in the Einstein gravity~\cite{Emparan:2014jca,Emparan:2016sjk}, 
the large $D$ limit is incompatible to the extremal limit, so that we need a some remedy to eliminate the apparent breaks down of the $1/n$ expansion near the extremal limit, as actually performed for charged squashed black holes~\cite{Suzuki:2021lrw}. 
%The large $D$ limit has been quite incompatible to the extremal limit already in the Einstein gravity~\cite{Emparan:2014jca,Emparan:2016sjk}.
%We need a remedy to absorb the apparent breaks down of the $1/n$ expansion near the extremal limit as was done in the case of charged squashed black holes~\cite{Suzuki:2021lrw}. 
Finding the analytic solution of equally rotating black holes in the pure GB theory could shed some light on the extremal limit in the EGB theory.
Interestingly, the extremal limit of the equally-rotating black holes was examined for small $\alpha$ in $D=5$~\cite{Ma:2020xwi}, where the inner horizon only appears close to the extremality.

\begin{figure}[h]
\begin{center}
\includegraphics[width=8cm]{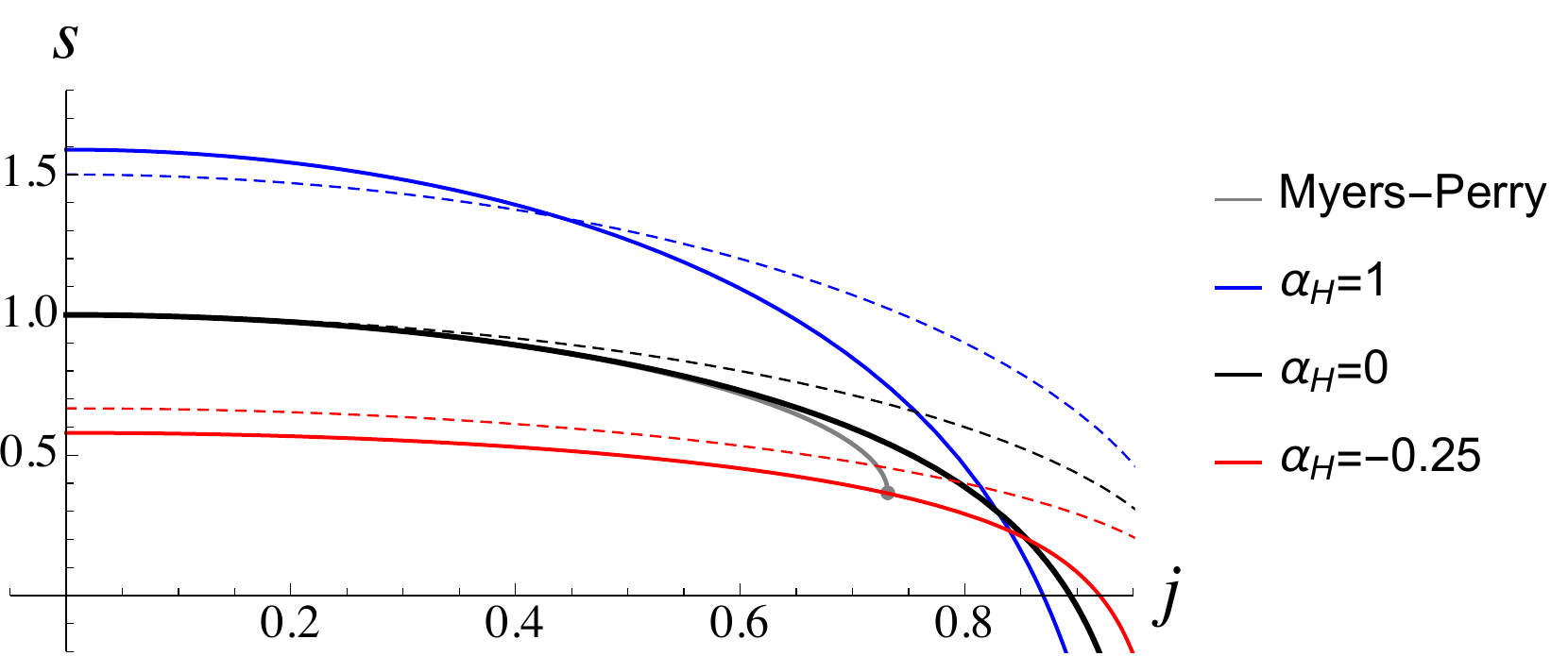}
\caption{The phase diagram in the space of the entropy and angular momentum normalized by the mass for $n=4 \,(D=11)$. The thick and dashed curves represents the NLO and LO results respectively.
The exact Myers-Perry solutions for $n=4$ are also plotted by the gray curve. \label{fig:sjplot}}
\end{center}
\end{figure}

Below, we present the phase diagram in the terms of dimensionless variables related by
%The phase diagram is presented in the angular momentum and entropy normalized by the mass (fig.~\ref{fig:sjplot})
\begin{align}
&s = \frac{\sqrt{1-j^2} \left(2 \alpha _H+1\right)}{\alpha _H+1}\nonum
& \hspace{1cm}\times \left[1+\fr{2n(1-j^2)}\left(\log\left(\frac{1-j^2}{1+\alpha_H}\right)\right.\right.\nonum
&\left.\left.\hspace{2cm}+\frac{\alpha _H \left(4 \left(1-j^2\right) \alpha _H-4 j^2+3\right)}{\left(\alpha
   _H+1\right) \left(2 \alpha _H+1\right)}\right)\right],
\end{align}
where the angular momentum and entropy are normalized by the mass scale
\begin{align}
& j:= \frac{8\pi G J}{(n+1)\Omega_{2n+1}} 
\left( \frac{8\pi G M}{(n+1/2)\Omega_{2n+1}}\right)^{-\frac{2n+1}{2n}},\\
& s:= \frac{4G S}{(n+1)\Omega_{2n+1}} 
\left( \frac{8\pi G M}{(n+1/2)\Omega_{2n+1}}\right)^{-\frac{2n+1}{2n}}.
\end{align}
Here the spin parameter is expressed as the function of $j$
\begin{align}
 \Omega = j - \frac{j}{2n}
 \left[\log\left(\frac{1-j^2}{1+\alpha_H}\right)-\frac{2\alpha_H j^2}{(1+\alpha_H)(1+2\alpha_H)} \right].
\end{align}
In fig.~\ref{fig:sjplot},
the phase diagram shows that the positive (negative) value of $\alpha$ gives larger (smaller) entropy than GR solutions  for each $j$ , succeeding the property of the static solutions. Near the extremality the convergence of $1/n$ expansion becomes bad.

\medskip
In this work, using the large $D$ approach, we have obtained the first analytic solutions of not-slowly rotating black holes to the EGB theory in odd dimensions.
For larger $\alpha$, the size of the ergoregion becomes larger, and then for $\alpha \to \infty$, it saturates.
We have also determined the first phase diagram of equally-rotating EGB black holes.
More technical details and higher order corrections will be presented in the forthcoming paper~\cite{STnext}.

%In the pure GB theory, it was shown that stable circular orbits exists in the static or singly rotating black holes for $D=6,7,8$, 

\medskip
By introducing the dependence of the metric on the time and angular coordinates, one can obtain the large $D$ effective theory which enables the stability analysis of the horizon.
%\red{???}We expect the same strategy also applies to the singly rotating case, which may additionally require solving the horizon embedding $r=r_0(\theta)$. 
We expect the same strategy also applies to the singly-rotating case.
It would be also interesting to explore the large $D$ rotating black holes in the more general Lovelock theory~\cite{Garraffo:2008hu} with the same ansatz.

\begin{acknowledgements}
RS was supported by JSPS KAKENHI Grant Number~JP18K13541.
ST was supported by JSPS KAKENHI Grant Number~17K05452 and 21K03560.
%ST was supported by the Grant-in-Aid for Scientific Research (C) [JSPS KAKENHI Grant Number~17K05452] and the Grant-in-Aid for Scientific Research (C) [JSPS KAKENHI Grant Number~21K03560] from the Japan Society for the Promotion of Science. 
\end{acknowledgements}


\begin{thebibliography}{99}


%\cite{Gross:1986iv}
\bibitem{Gross:1986iv}
D.~J.~Gross and E.~Witten,
%``Superstring Modifications of Einstein's Equations,''
Nucl. Phys. B \textbf{277}, 1 (1986).
%doi:10.1016/0550-3213(86)90429-3
%890 citations counted in INSPIRE as of 20 Feb 2022


%\cite{Antoniadis:1997eg}
\bibitem{Antoniadis:1997eg}
I.~Antoniadis, S.~Ferrara, R.~Minasian and K.~S.~Narain,
%``R**4 couplings in M and type II theories on Calabi-Yau spaces,''
Nucl. Phys. B \textbf{507}, 571-588 (1997)
%doi:10.1016/S0550-3213(97)00572-5
[arXiv:hep-th/9707013 [hep-th]].
%209 citations counted in INSPIRE as of 20 Feb 2022


%\cite{Ferrara:1996hh}
\bibitem{Ferrara:1996hh}
S.~Ferrara, R.~R.~Khuri and R.~Minasian,
%``M theory on a Calabi-Yau manifold,''
Phys. Lett. B \textbf{375}, 81-88 (1996)
%doi:10.1016/0370-2693(96)00270-5
[arXiv:hep-th/9602102 [hep-th]].

%Gauss-Bonnet theory and black holes

%\cite{Boulware:1985wk}
\bibitem{Boulware:1985wk}
D.~G.~Boulware and S.~Deser,
%``String Generated Gravity Models,''
Phys. Rev. Lett. \textbf{55}, 2656 (1985).
%doi:10.1103/PhysRevLett.55.2656
%1197 citations counted in INSPIRE as of 01 Feb 2022



%\cite{Wiltshire:1988uq}
\bibitem{Wiltshire:1988uq}
D.~L.~Wiltshire,
%``Black Holes in String Generated Gravity Models,''
Phys. Rev. D \textbf{38}, 2445 (1988).
%doi:10.1103/PhysRevD.38.2445
%280 citations counted in INSPIRE as of 07 Feb 2022

%\cite{Wiltshire:1985us}
\bibitem{Wiltshire:1985us}
D.~L.~Wiltshire,
%``Spherically Symmetric Solutions of Einstein-maxwell Theory With a {Gauss-Bonnet} Term,''
Phys. Lett. B \textbf{169}, 36-40 (1986).
%doi:10.1016/0370-2693(86)90681-7
%302 citations counted in INSPIRE as of 07 Feb 2022

%\cite{Brihaye:2008kh}
\bibitem{Brihaye:2008kh}
Y.~Brihaye and E.~Radu,
%``Five-dimensional rotating black holes in Einstein-Gauss-Bonnet theory,''
Phys. Lett. B \textbf{661}, 167-174 (2008)
%doi:10.1016/j.physletb.2008.02.005
[arXiv:0801.1021 [hep-th]].
%57 citations counted in INSPIRE as of 07 Feb 2022


%\cite{Kim:2007iw}
\bibitem{Kim:2007iw}
H.~C.~Kim and R.~G.~Cai,
%``Slowly Rotating Charged Gauss-Bonnet Black holes in AdS Spaces,''
Phys. Rev. D \textbf{77}, 024045 (2008)
%doi:10.1103/PhysRevD.77.024045
[arXiv:0711.0885 [hep-th]].
%66 citations counted in INSPIRE as of 07 Feb 2022


% Large D

%\cite{Asnin:2007rw}
\bibitem{Asnin:2007rw}
V.~Asnin, D.~Gorbonos, S.~Hadar, B.~Kol, M.~Levi and U.~Miyamoto,
%``High and Low Dimensions in The Black Hole Negative Mode,''
Class. Quant. Grav. \textbf{24}, 5527-5540 (2007)
%doi:10.1088/0264-9381/24/22/015
[arXiv:0706.1555 [hep-th]].
%58 citations counted in INSPIRE as of 31 Jan 2022

%\cite{Emparan:2013moa}
\bibitem{Emparan:2013moa}
R.~Emparan, R.~Suzuki and K.~Tanabe,
%``The large D limit of General Relativity,''
JHEP \textbf{06}, 009 (2013)
%doi:10.1007/JHEP06(2013)009
[arXiv:1302.6382 [hep-th]].
%173 citations counted in INSPIRE as of 31 Jan 2022



%\cite{Emparan:2020inr}
\bibitem{Emparan:2020inr}
R.~Emparan and C.~P.~Herzog,
%``Large D limit of Einstein\textquoteright{}s equations,''
Rev. Mod. Phys. \textbf{92}, no.4, 045005 (2020)
%doi:10.1103/RevModPhys.92.045005
[arXiv:2003.11394 [hep-th]].
%22 citations counted in INSPIRE as of 31 Jan 2022


%\cite{Emparan:2015hwa}
\bibitem{Emparan:2015hwa}
R.~Emparan, T.~Shiromizu, R.~Suzuki, K.~Tanabe and T.~Tanaka,
%``Effective theory of Black Holes in the 1/D expansion,''
JHEP \textbf{06} (2015), 159
%doi:10.1007/JHEP06(2015)159
[arXiv:1504.06489 [hep-th]].


%\cite{Bhattacharyya:2015dva}
\bibitem{Bhattacharyya:2015dva}
S.~Bhattacharyya, A.~De, S.~Minwalla, R.~Mohan and A.~Saha,
%``A membrane paradigm at large D,''
JHEP \textbf{04} (2016), 076
%doi:10.1007/JHEP04(2016)076
[arXiv:1504.06613 [hep-th]].

%\cite{Bhattacharyya:2015fdk}
\bibitem{Bhattacharyya:2015fdk}
S.~Bhattacharyya, M.~Mandlik, S.~Minwalla and S.~Thakur,
%``A Charged Membrane Paradigm at Large D,''
JHEP \textbf{04} (2016), 128
%doi:10.1007/JHEP04(2016)128
[arXiv:1511.03432 [hep-th]].


%\cite{Emparan:2015gva}
\bibitem{Emparan:2015gva}
  R.~Emparan, R.~Suzuki and K.~Tanabe,
%  ``Evolution and End Point of the Black String Instability: Large D Solution,''
  Phys.\ Rev.\ Lett.\  {\bf 115} (2015) no.9,  091102
%  doi:10.1103/PhysRevLett.115.091102
  [arXiv:1506.06772 [hep-th]].
  %%CITATION = doi:10.1103/PhysRevLett.115.091102;%%
  
  % Large D and EGB theory

%\cite{Chen:2017hwm}
\bibitem{Chen:2017hwm}
B.~Chen and P.~C.~Li,
%``Static Gauss-Bonnet Black Holes at Large $D$,''
JHEP \textbf{05}, 025 (2017)
%doi:10.1007/JHEP05(2017)025
[arXiv:1703.06381 [hep-th]].
%28 citations counted in INSPIRE as of 31 Jan 2022

%\cite{Chen:2017rxa}
\bibitem{Chen:2017rxa}
B.~Chen, P.~C.~Li and C.~Y.~Zhang,
%``Einstein-Gauss-Bonnet Black Strings at Large $D$,''
JHEP \textbf{10}, 123 (2017)
%doi:10.1007/JHEP10(2017)123
[arXiv:1707.09766 [hep-th]].
%27 citations counted in INSPIRE as of 31 Jan 2022


%\cite{Chen:2018vbv}
\bibitem{Chen:2018vbv}
B.~Chen, P.~C.~Li and C.~Y.~Zhang,
%``Einstein-Gauss-Bonnet Black Rings at Large $D$,''
JHEP \textbf{07}, 067 (2018)
%doi:10.1007/JHEP07(2018)067
[arXiv:1805.03345 [hep-th]].
%11 citations counted in INSPIRE as of 31 Jan 2022  

%\cite{Emparan:2013xia}
\bibitem{Emparan:2013xia}
R.~Emparan, D.~Grumiller and K.~Tanabe,
%``Large-D gravity and low-D strings,''
Phys. Rev. Lett. \textbf{110}, no.25, 251102 (2013)
%doi:10.1103/PhysRevLett.110.251102
[arXiv:1303.1995 [hep-th]].
%88 citations counted in INSPIRE as of 31 Jan 2022
%\cite{Emparan:2014jca}
\bibitem{Emparan:2014jca}
R.~Emparan, R.~Suzuki and K.~Tanabe,
%``Instability of rotating black holes: large D analysis,''
JHEP \textbf{06}, 106 (2014)
%doi:10.1007/JHEP06(2014)106
[arXiv:1402.6215 [hep-th]].
%56 citations counted in INSPIRE as of 31 Jan 2022


\bibitem{Mandlik:2018wnw}
M.~Mandlik and S.~Thakur,
%``Stationary Solutions from the Large D Membrane Paradigm,''
JHEP \textbf{11}, 026 (2018)
%doi:10.1007/JHEP11(2018)026
[arXiv:1806.04637 [hep-th]].
%15 citations counted in INSPIRE as of 31 Jan 2022

%\cite{Tanabe:2016opw}
\bibitem{Tanabe:2016opw}
K.~Tanabe,
%``Charged rotating black holes at large D,''
[arXiv:1605.08854 [hep-th]].
%27 citations counted in INSPIRE as of 31 Jan 2022





\bibitem{Wald:1993nt}
R.~M.~Wald,
%``Black hole entropy is the Noether charge,''
Phys. Rev. D \textbf{48}, no.8, R3427-R3431 (1993)
%doi:10.1103/PhysRevD.48.R3427
[arXiv:gr-qc/9307038 [gr-qc]].
%1842 citations counted in INSPIRE as of 01 Feb 2022
\bibitem{Iyer:1994ys}
V.~Iyer and R.~M.~Wald,
%``Some properties of Noether charge and a proposal for dynamical black hole entropy,''
Phys. Rev. D \textbf{50}, 846-864 (1994)
%doi:10.1103/PhysRevD.50.846
[arXiv:gr-qc/9403028 [gr-qc]].
%1608 citations counted in INSPIRE as of 01 Feb 2022

\bibitem{Suzuki:2021lrw}
R.~Suzuki and S.~Tomizawa,
%``Squashed black holes at large D,''
JHEP \textbf{12}, 194 (2021)
%doi:10.1007/JHEP12(2021)194
[arXiv:2111.04962 [hep-th]].
%0 citations counted in INSPIRE as of 04 Feb 2022


%\cite{Emparan:2016sjk}
\bibitem{Emparan:2016sjk}
  R.~Emparan, K.~Izumi, R.~Luna, R.~Suzuki and K.~Tanabe,
%  ``Hydro-elastic Complementarity in Black Branes at large D,''
  JHEP {\bf 1606} (2016) 117
%  doi:10.1007/JHEP06(2016)117
  [arXiv:1602.05752 [hep-th]].
  %%CITATION = doi:10.1007/JHEP06(2016)117;%%


%\cite{Ma:2020xwi}
\bibitem{Ma:2020xwi}
L.~Ma, Y.~Z.~Li and H.~Lu,
%``D = 5 rotating black holes in Einstein-Gauss-Bonnet gravity: mass and angular momentum in extremality,''
JHEP \textbf{01}, 201 (2021)
%doi:10.1007/JHEP01(2021)201
[arXiv:2009.00015 [hep-th]].
%2 citations counted in INSPIRE as of 05 Feb 2022

\bibitem{STnext}
R.~Suzuki and S.~Tomizawa, to appear.

\bibitem{Garraffo:2008hu}
C.~Garraffo and G.~Giribet,
%``The Lovelock Black Holes,''
Mod. Phys. Lett. A \textbf{23}, 1801-1818 (2008)
%doi:10.1142/S0217732308027497
[arXiv:0805.3575 [gr-qc]].
%155 citations counted in INSPIRE as of 01 Feb 2022


\end{thebibliography}
\end{document}